\documentclass[prl,twocolumn,preprintnumbers,amsmath,amssymb,superscriptaddress,longbibliography]{revtex4-2}
\usepackage{xcolor}
\usepackage{natbib}
\usepackage{svg}
\usepackage{appendix}
\usepackage{amsfonts,amsmath,amsthm}
\usepackage{bm}
\usepackage{graphicx}
\usepackage{dcolumn}
\usepackage{braket}    
\usepackage{physics}   
\usepackage{bm}
\def\theYear{\the\year}

\begin{document}

\setcounter{page}{1}


\title{Phase assumption-free multiparty quantum clock synchronization}

\author{Hatim A. Oujaa}
\affiliation{New York University Shanghai; NYU-ECNU Institute of Physics at NYU Shanghai, 567 West Yangsi Road, Shanghai, 200124, China.}
\affiliation{Ecole Normale Supérieure Paris-Saclay, 4 Av. des Sciences, 91190 Gif-sur-Yvette, France}
\affiliation{CentraleSupélec, 3 Rue Joliot Curie, 91190 Gif-sur-Yvette, France}

\author{Qiao Liu}
\affiliation{Department of Materials Science and Engineering; School of Applied and Engineering Physics, Cornell University, Ithaca, NY 14853, USA}

\author{Ebubechukwu O. Ilo-Okeke}
\affiliation{Department of Physics, College of William and Mary, Williamsburg, Virginia 23187, USA}

\author{Valentin Ivannikov}
\email{valentin@nyu.edu}
\affiliation{New York University Shanghai; NYU-ECNU Institute of Physics at NYU Shanghai, 567 West Yangsi Road, Shanghai, 200124, China.}

\author{Jonathan P. Dowling}
\affiliation{Hearne Institute for Theoretical Physics and Department of Physics and Astronomy, Louisiana State University, Baton Rouge, Louisiana 70803, USA}

\author{Tim Byrnes}
\email{tim.byrnes@nyu.edu}
\affiliation{New York University Shanghai; NYU-ECNU Institute of Physics at NYU Shanghai, 567 West Yangsi Road, Shanghai, 200124, China.}
\affiliation{State Key Laboratory of Precision Spectroscopy, School of Physical and Material Sciences, East China Normal University, Shanghai 200062, China}
\affiliation{Center for Quantum and Topological Systems (CQTS), NYUAD Research Institute, New York University Abu Dhabi, UAE.}
\affiliation{Department of Physics, New York University, New York, NY 10003, USA}

\date{\today}

\begin{abstract}
We investigate methods to broadcast timing information from a central clock to all other clocks by the use of multipartite entanglement.  This task is a necessary step in establishing a coordinated universal time, currently performed using classical synchronization methods.  Using an entanglement-based method has the advantage that the timing results are independent of the intervening medium.  We generalize existing bipartite quantum clock synchronization methods and take special care to address issues of different phase conventions being adopted at each node (the ``Preskill phase problem'').  Using supersinglet purification, we show that this allows for a scalable method with a time signal that is a constant with respect to the number of nodes.  
\end{abstract}

\maketitle

\paragraph{Introduction}

The goal of obtaining a universal standard time, where a common time is established across numerous parties around the world, is a crucial technology in modern day society \cite{panfilo2016coordinated,panfilo2019coordinated,arias2004coordinated}. Among numerous applications \cite{zhou2016applications}, those which require high accuracy include sensor networks \cite{giovannetti2001quantum}, global positioning system satellites \cite{enge1994global}, and gravitational wave observation \cite{aasi2015advanced}.  One of the main tasks in achieving  a  coordinated universal time (UTC) is to synchronize accurate clocks. Currently, UTC is obtained by taking the weighted average of atomic clocks around the world, which is then used as the standard for all clocks to synchronize against. The synchronization is typically performed using Two Way Satellite Time and Frequency Transfer (TWSTFT) \cite{kirchner1999two,lewandowski1991gps}, which involves two-way radio communication via a geostationary satellite. The primary reason that the two-way communication is used is due to cancel out delays due to fluctuations in the atmosphere and satellite transponder delays.  The time-transfer accuracy of TWSTFT is typically at the 1 ns level \cite{zhang2009two,jiang2019improving}.


Quantum clock synchronization (QCS) is an alternative method for clock synchronization that has emerged with advantages over classical methods.  First introduced by Jozsa, Abrams, Dowling, and Williams \cite{jozsa2000quantum}, the method attains synchronization between two distant parties by the use of qubits in entangled states.  The qubits represent precisely energy separated levels, which can constitute the clock states in an atomic clock, for example.  Then by one party performing a measurement, the precession of the qubit  in the remote clock is initiated, thereby remotely sharing timing information between two parties.  The QCS protocol has the potential for improving the accuracy of clock synchronization because no actual timing information needs to be communicated between the parties, only the measurement result of the transmitter \cite{giovannetti2004quantum}.  Picosecond precision should be attainable given that there are a sufficient number of shared entangled qubits \cite{ilo2018remote}.   One criticism of the QCS protocol made by Preskill shortly after its proposal was that there is a hidden assumption of a common phase reference when defining the original entangled state \cite{preskill2000}.  It was pointed out that to have a common phase reference, synchronized clocks would be required, defeating the purpose of QCS.  This issue ---- dubbed the ``Preskill phase problem'' --- was subsequently resolved in the two-party case by performing a preliminary step of performing entanglement purification, which simultaneously prepares the entangled state, and establishes a common phase reference \cite{ilo2018remote}.  Several experiments have been done to test the QCS protocol \cite{zhang2004,valencia2004distant,quan2016,kong2017,kong2017implementation}. Alternative methods based on different principles for synchronization to the QCS protocol have also been suggested \cite{chuang2000quantum,giovannetti2001clock,giovannetti2002positioning,de2005quantum,tavakoli2015quantum}.  

Several approaches to generalize the original two-party QCS protocol to the multiparty case have been performed.  This would be directly applicable to sychronization tasks such as establishing UTC, for example, due to the numerous clocks stationed around the world. In the approach of Ref.  \cite{krco2002}, a W state is shared between the parties, and Alice shares timing information by performing a measurement.  Other parties make measurements to extract the time signal with the help of classical communication.  The approach has the disadvantage that the signal degrades as $ \propto 1/N $, where $ N $ is the number of parties.  An improvement upon this was made in Ref.  \cite{ben2011optimized}, where Dicke states are used as the initial entangled state.  Using the central Dicke state, the signal scales as a constant with respect to $ N $. Dicke states (including W states) are advantageous in the task of multiparty QCS since the entanglement is of a bipartite nature, rather than genuinely multipartite.  Multiparty schemes using GHZ states have a non-local signal in the sense that all parties must work together to extract timing information \cite{ren2012clock,komar2014quantum}, and hence is less desirable in the context of synchronization.   
In all these works, the Preskill phase problem has not been addressed, which is a critical component of a realistic protocol where phase conventions are independent.  Without resolving this issue, the timing information has a systematic offset, rendering the QCS protocol ineffective.

\begin{figure}[t]   
\includegraphics[width=\linewidth]{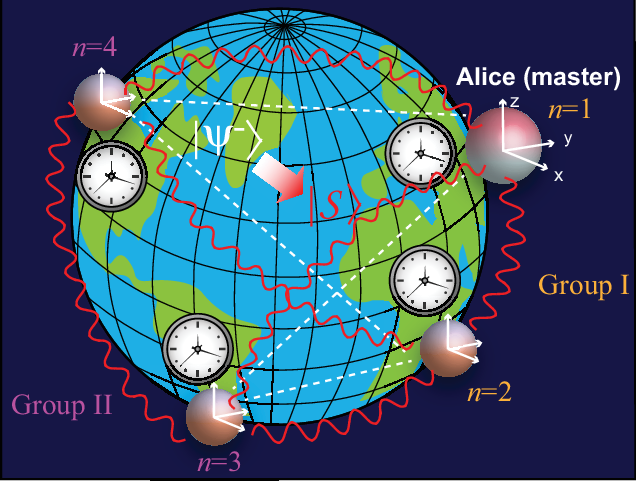}
\caption{Quantum time broadcasting using supersinglet states. Alice wishes to share her timing information with the remaining parties, each holding a qubit depicted by a Bloch sphere (we show the $ N = 4 $ case).  The full protocol proceeds as follows. Step 1: Distribute singlet pairs $ |\Psi^- \rangle $ between parties following symmetry of supersinglet state (dashed lines).  Interchange of qubits within Group I ($ n \in [1,N/2] $) and Group II ($ n \in [N/2+1,N] $) leaves the distribution symmetric.  Step 2: Perform entanglement purification following Ref. \cite{bennett1996} or similar.  Step 3: Check correlation of $ \langle  Z_i Z_j  \rangle $ of singlet pairs, if not consistent with singlet state, then apply a $ \pi $ rotation around $ Z $ axis and repeat Step 2. At this point the Preskill phase is removed. Step 4: Perform  purification protocol in Ref. \cite{ahmad2025distillation} to obtain a supersinglet state $ | {\cal S} \rangle $ (wavy lines).  Step 6: Alice measures her state in the $ X $ basis and classically broadcasts the result. Step 7: Each party measures their local qubit in the $ X $ basis, and adjusts the amplitude according to Alice's outcome.  Step 8: Parties infer the time of Alice's measurement using (\ref{qcsamp}) and (\ref{supersingletamplitudes}).  }
\label{fig1}
\end{figure}

In this paper, we develop a full protocol for the task of broadcasting timing information using shared entanglement.  The protocol is summarized in Fig. \ref{fig1}.  Alice is the central (master) clock from which time information needs to be distributed to all other clocks.  In the state initialization stage, bipartite singlet pairs are distributed, which are converted to a supersinglet state using purification \cite{ahmad2025distillation}. The supersinglet state, introduced by Cabello \cite{cabello2003supersinglets}, is a particular type of spin-zero state that has a non-trivial entanglement structure that we show is well-suited to our current task.  In particular, we show the purification resolves the Preskill phase problem, which is crucial for long-distance applications.  We show that the supersinglet state can then be used for multiparty QCS, generalizing the bipartite case.  
We show that the supersinglet state has good scaling properties of the timing signal with respect to the number of parties $ N $, making it suitable for applications with a large number of clocks to be synchronized.   

\paragraph{Singlet based QCS}

Consider $ N $ parties (taken to be even number throughout this paper), each in possession of an accurate clock. The frequency of the clocks can be considered to be accurate, but their relative phases are unsychronized.  
Each party possesses a qubit with precisely separated energy levels with the Hamiltonian
%
\begin{align}
H = \frac{\hbar \omega}{2} \sum_{n=1}^{N} Z_n = \hbar \omega S^z
\label{ham}
\end{align}
where the Pauli matrix $ Z_n = |0 \rangle \langle 0 |_n - |1 \rangle \langle 1 |_n $ for the $ n $th qubit.  Here we defined the total spin operator $ S^z = \sum_{n=1}^{N} Z_n/2 $, and similarly defined for the $ x,y $ axes.  The total spin eigenstates are defined as $ S^2 | s,\alpha ,m \rangle^{(z)} = s(s+1) | s,\alpha ,m \rangle^{(z)} $, $ S^z | s,\alpha ,m \rangle^{(z)} = m | s,\alpha ,m \rangle^{(z)} $, where $ m \in [-s,s] $, and $ S^2 = (S^x)^2 + (S^y)^2 + (S^z)^2$.  The $ ^{(z)} $ denotes that the eigenvalues 
are for the $ S^z $ operator.  The label $ \alpha  $ is the outer multiplicity label. 

The $ N $ parties are initially in possession of a singlet state  $ | {\cal S} \rangle = \sum_\alpha \phi_\alpha | s = 0,\alpha ,m=0 \rangle $ satisfying 
\begin{align}
S^2  | {\cal S} \rangle = S^j | {\cal S} \rangle =  0 ,
\end{align}
for any basis $ j \in \{x,y,z\} $.  For any $ N \ge 4  $, the singlet sector has multiplicity and there is a choice of the type of singlet state, giving rise to the coefficients $ \phi_\alpha $ (to be determined below). 
The state $ | {\cal S} \rangle  $ has a zero eigenvalue with respect to the Hamiltonian (\ref{ham}), and hence is unaffected by its time evolution.  Any singlet state can be decomposed as
\begin{align}
| {\cal S} \rangle = & \frac{1}{\sqrt{2}} \Big[ |+ \rangle_1 |s=1/2, m=-1/2 \rangle_{N-1}^{(x)} \nonumber \\
& - |-  \rangle_1 |s=1/2,  m =1/2 \rangle_{N-1}^{(x)}  \Big] ,
\label{belllike}
\end{align}
where $ | \pm \rangle = (|0 \rangle \pm |1 \rangle )/\sqrt{2}  $ and $ |s=1/2,m \rangle_{N-1}^{(x)} $ is a $ N -1 $  qubit state forming a $ s=1/2 $ state via angular momentum coupling.  Therefore, after Alice (in possession of the $ n =1$ qubit) measures her qubit in the $ x $ basis, the state collapses to $ |\pm \rangle_1 |s=1/2,m= \mp 1/2 \rangle_{N-1}^{(x)} $. Alice announces her result to all the parties.  Since Alice's qubit is decoupled from the remaining system, we drop it henceforth. We note that the states $ |s=1/2,m= \mp 1/2 \rangle_{N-1}^{(x)} $ are not unique since it depends upon the particular initial singlet state $ | {\cal S} \rangle $, and for large $ N $ there are many ways that spins may coupled to form a total $ s = 1/2 $.  

Under time evolution of the Hamiltonian (\ref{ham}) the state is $ e^{-i H t/\hbar }  |s=1/2, m= \mp 1/2 \rangle_{N-1}^{(x)} $.
Crucially, the time $ t $ is the time elapsed {\it since} Alice's measurement is made, which is the basis of the QCS protocol. In order to obtain a deterministic result, for the case that Alice measures $ | - \rangle $, an additional rotation of the state $ \exp ( i \pi S^z ) = \prod_{n=1}^N \exp (i \pi Z_n /2) $ is performed by all parties (or alternatively can be accounted for by classical post-processing, see Appendix).  At this point we deterministically have the state
\begin{align}
|\psi(t) \rangle & = e^{-i H t/\hbar }  |s=1/2, m=  -1/2 \rangle_{N-1}^{(x)} 
\label{postmeasuredpsi}
\end{align}
on the qubits $ n \in [2,N] $.  
The expectation value of the $ n $th qubit may be evaluated to be
\begin{align}
f_n(t) &  = \langle \psi(t) |  X_n | \psi(t) \rangle = A_n \cos \omega t  \label{qcsamp} .
\end{align}
%
where $ A_n  = \langle \psi(0) | X_n | \psi(0) \rangle $.  From the fact that the state $ | \psi (0) \rangle $ on qubits $ n \in [2,N ] $ forms a $ s = 1/2 $ state, we also have
\begin{align}
\langle \psi(0)  |  \sum_{n=2}^N X_n | \psi(0)  \rangle=   \sum_{n=2}^{N} A_n = -1 .
\label{totalamp}
\end{align}

The signal (\ref{qcsamp}) for each of party $ n \in [2,N] $ can be obtained with a local measurement, and no communication beyond Alice's measurement outcome needs to be performed, which is an advantage of over GHZ state based methods which must still share information.  We also note that the measurements on different nodes all commute and do not affect measurement results on other parties. Thus the precise ordering or timing of the measurements on qubits to obtain $ \langle X_n \rangle $ do not matter. Repeating the measurement on an ensemble of singlet states (\ref{belllike}) and taking statistics in the usual way recovers the signal (\ref{qcsamp}).

\paragraph{Choice of singlet state}

We find that the optimal type of singlet state that gives the largest magnitude amplitude of the signals for the parties is the supersinglet state \cite{cabello2003supersinglets}, defined as 
\begin{align}
| {\cal S}_{N} \rangle =\sum_{k=0}^{N/2}  \frac{(-1)^k}{\sqrt{N/2+1}}  
|D_k^{N/2} \rangle |D_{N/2-k}^{N/2} \rangle  .
\label{supersingletdef}
\end{align}
where $ |D_k^{N/2} \rangle = |s=N/4,1,m= N/4-k\rangle_{N/2}^{(z)} $ are the Dicke states constructed from $N/2 $ qubits \cite{byrnes2021quantum}.  This type of singlet state is constructed by grouping the qubits into two groups of $ N/2 $ qubits, each forming a maximum spin state $ s = N/4 $, then these are combined to form a total spin zero state.  We evaluate that the amplitudes of the signals (\ref{qcsamp}) when using the supersinglet state are (see Appendix)
\begin{align}
A_n = \left\{
\begin{array}{cc}
    1/3 & n \in [2,N/2] \\
    -\frac{N+4}{3N} & n \in [N/2+1,N]
\end{array}
\right. .
\label{supersingletamplitudes}
\end{align}
The difference in the amplitudes arise from the fact that Alice performs a measurement on the first qubit, which is in the first group of qubits, creating the asymmetry. For large $ N $, asymptotically both groups have an amplitude with magnitude $ 1/3$.  
Other choices of singlet states have a smaller magnitude of oscillations (see Appendix).  

Due to the inhomogenous amplitudes (\ref{supersingletamplitudes}), we require an additional step in the context of the QCS protocol. In the initialization stage, each party is notified whether they are in Group I (for qubits $ n \in [1,N/2] $) or Group II (for qubits $ n \in [N/2+1,N] $). Then all signals are normalized by the amplitudes given in (\ref{supersingletamplitudes}), such that a consistent signal is obtained.  This simple modification ensures that symmetry is restored across the network, and enabling consistent synchronization across all parties. 

\begin{figure}[t]   
\includegraphics[width=\linewidth]{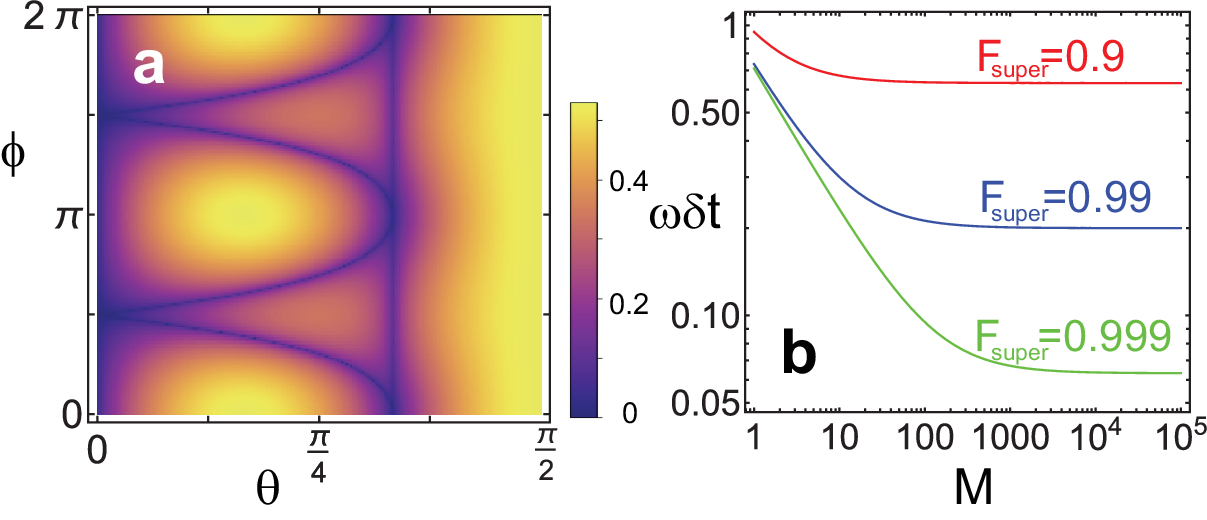}
\caption{Performance of the multiparty quantum clock synchronization.  (a) Optimization with respect to singlet states for the $ N = 4$ case using parametrization $  \ket{\psi} = \cos(\theta)\ket{\Psi^-}_{12} \ket{\Psi^-}_{34}  + e^{i\phi}\sin(\theta)\ket{\mathcal{S}_4} $.   As the function to be optimized, we take $ \left( \prod_{n=2}^{N} |A_n | \right)^{\frac{1}{N-1}} $, the geometric mean of the QCS amplitudes. (b) Error estimate (\ref{totalerror}) of multiparty QCS for various supersinglet fidelities $ F_{\text{super}} $.  
}
\label{fig2}
\end{figure}

The supersinglet state is optimal for the following reasons.  The range of values the amplitude can take are $ -1 \le A_n \le 1/3 $ for any spin zero state, due to the restriction of SU(2) invariance (see Appendix).  Under the constraint (\ref{totalamp}), it is beneficial to take as many $ A_n $ to be 1/3 as possible to maximize the amplitudes $ |A_n| $.  This however occurs only for a ferromagnetic alignment of qubits, where the spin is $ s = M/2 $ for $ M $ qubits.  Then one must maximize $ M $, while also finally resulting in a total spin zero state, which is optimall attained with the supersinglet state (\ref{supersingletdef}) for $ M = N/2 $.  To numerically verify the optimality, we directly optimize the singlet states for the $ N = 4$ case (Fig. \ref{fig2}(a)).   As expected, we obtain the optimal point as the supersinglet state, corresponding to $\theta = \pi/2$. The other two states, corresponding to ($\theta = \pi/6$,  $\phi=0, \pi$), are equivalent supersinglet states, up to permutations of parties between the two groups.

\paragraph{Preskill phase} 
In order to obtain a genuine multiparty QCS protocol, it must only use only local operations and classical communications (LOCC) and not make hidden phase assumptions.   The Preskill phase problem \cite{preskill2000quantum} can be summarized for QCS as follows. We assume that the clocks are separated by large (e.g. intercontinental) distances.  While the qubit states $ \{ | 0 \rangle_n, |1 \rangle_n \} $ at each location $ n $ are well-defined by their physical properties (e.g. atomic levels), their superpositions are only defined by convention.  For example, the state $ | + \rangle = (| 0 \rangle +| 1 \rangle)/\sqrt{2} $ will appear different if one uses a different phase convention for the logical states $ | 0' \rangle = e^{i \theta_0} | 0 \rangle, | 1' \rangle = e^{i \theta_1} | 1 \rangle $.  This phase ambiguity will affect the QCS protocol since this corresponds to an uncontrolled systematic $Z_n $ rotation.  In our context, this will produce a timing offset in (\ref{qcsamp}). 

The key to resolving this issue is to {\it prepare the supersinglet state using the same phase conventions} as the measurements in the QCS protocol. If the supersinglet (or any other entangled state) is produced in a different laboratory and  distributed to the remote sites, it is susceptible to the Preskill phase because the phase conventions may not match \cite{chaudhary2023macroscopic,mao2023measurement}.  There is an obvious difficulty here, since an entangled state must be prepared on distant qubits, which can only implement LOCC.  Entanglement purification is ideally suited to solving this problem, since it allows for the preparation of an entangled state using only LOCC.  Specifically, the step of performing bilateral rotations (``twirling'') \cite{bennett1996} has the singlet state as a fixed point, {\it with respect to the local basis} \cite{ilo2018remote}.  

\begin{figure}[t]   
\includegraphics[width=\linewidth]{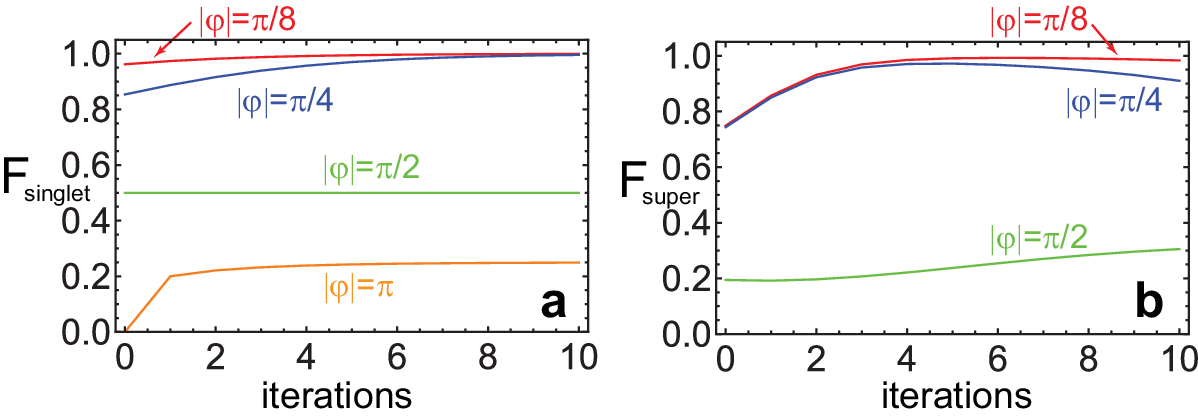}
\caption{Entanglement purification of supersinglet states with the Preskill phase. Fidelities are defined as $ F_{\text{singlet}} = \langle \Psi^- | \rho | \Psi^- \rangle $ and $ F_{\text{super}} = \langle {\cal S} | \rho | {\cal S} \rangle $. We target the $ N = 4 $ supersinglet.   Purification iterations are shown for (a) singlet Bell pairs using BBPSSW purification \cite{bennett1996}; (b) supersinglets using methods in Ref. \cite{ahmad2025distillation}.  The initial state for the singlet Bell pair purification of (a) is $ \exp (i \varphi Z_1/2) | \Psi^- \rangle $, where $ \varphi $ is the Preskill phase.  The output of BBPSSW purification after 10 rounds is used as the input for the supersinglet purification in (b).   }
\label{fig3}
\end{figure}

To prepare the supersinglet state, we use the protocol introduced in Ref. \cite{ahmad2025distillation}. In this protocol, first singlet Bell states $ |\Psi^- \rangle $  are prepared using conventional entanglement purification techniques such as Ref. \cite{bennett1996}.  At this stage, as already shown in Ref. \cite{ilo2018remote}, the states will be singlet states defined in the local basis. In Fig. \ref{fig3}(a) we show how entanglement purification overcomes an uncontrolled Preskill phase $ \varphi $, as long as  $ |\varphi |< \pi/2 $.  Since the Preskill phase is a systematic error, one may correct for cases with $ |\varphi |\ge  \pi/2 $ by checking the correlation $ \langle Z_1 Z_2 \rangle $ on a sample of the post-purified states which converges to 0 rather than $-1$ for a successful purification.  Once this is established, application of $ e^{i \pi Z_1/2} $ puts the states in the correct sector.  
Henceforth, the distillation proceeds as normal as given in Ref. \cite{ahmad2025distillation}, since all operations are LOCC, and all qubits have the same local phase definition.  In this way, the supersinglet state (\ref{supersingletdef}) in the local phase basis convention can be created. In Fig. \ref{fig3}(b) we show the convergence of the supersinglet states with imperfect singlet state convergence. The protocol of Ref. \cite{ahmad2025distillation} is known to degrade for large number of iterations with a population outside the $ s = 0 $ sector.  Even for imperfect singlet state, high fidelity supersinglet states can be obtained after $ \sim 5 $ iterations.

\paragraph{Error estimates}

To assess the accuracy of the our multiparty QCS scheme, we evaluate the error of signal (\ref{qcsamp}).  We examine two dominant sources of error.  First, imperfect fidelity $ F_{\text{super}} $ of the supersinglet state due to insufficient rounds of purification (Fig. \ref{fig3}), decoherence (see Appendix), or otherwise, can lead to a loss accuracy of the QCS signal. Second, the signal (\ref{qcsamp}) is obtained by quantum measurements which suffer from shot noise.  This is reduced by preparing a sufficiently large number of copies $ M $ of the supersinglet state.  We evaluate that in our case the total error of the timing signal of these effects is (see Appendix)
\begin{align}
\delta t & \approx \frac{1}{\omega} \sqrt{ \frac{1}{2M} + 4( 1-F_{\text{super}}) } .
\label{totalerror}
\end{align}
Estimates for the error are plotted in Fig. \ref{fig2}(b).  The error is plotted in dimensionless units, some typical values of clock frequencies are $ \omega^{-1} = 17 $ ps for Cs or $ \omega^{-1} = 0.4 $ fs for Sr.  
We see that the errors improve consistently with $ M $, until they are saturated by the imperfect fidelity.  Even with modest $ M \approx 10 $ and fidelities $ F \approx 0.99 $, errors in the range of $ \delta t \sim 10 $ ps can be attained with Cs, which exceeds the best accuracies by TWSTFT \cite{zhang2009two,jiang2019improving}.

\paragraph{Conclusion}

We have introduced a multiparty clock distribution protocol based on supersinglet states, including the state preparation stage which overcomes the Preskill phase problem.  The singlet based scheme is a natural generalization of the two-party QCS protocol, and has comparable performance with alternative Dicke state based methods \cite{ben2011optimized}.  The primary advantage is that supersinglets have a LOCC purification protocol \cite{ahmad2025distillation} which is absent for general Dicke states.  This allows one to overcome the Preskill phase problem, which allows the QCS protocol to be used for long-distance applications, where a common phase reference is absent. The accuracy that is attainable is limited by the supersinglet fidelity and shot noise but performance outperforming classical methods should be possible with reasonable experimental parameters.  We consider the main experimental challenge to be the reliable generation of long-distance entanglement (likely initially generated by photons) to be transferred to matter qubits with a precise clock frequency according to (\ref{ham}).  While distribution of long-distance entanglement has been demonstrated using space-based methods \cite{yin2017}, its reliable transfer to matter qubits suffer from low success probabilities \cite{moehring2007entanglement,ritter2012elementary,stephenson2020high,saha2025high}.  One benefit of the current protocol is that the entanglement can be produced offline, so that the synchronization may proceed only when the entanglement is in place and suitably purified. 
Our protocol further motivates applications of a world-spanning quantum internet \cite{rohde2025quantum}.

\begin{acknowledgments}
This work is supported by the SMEC Scientific Research Innovation Project (2023ZKZD55); the Science and Technology Commission of Shanghai Municipality (22ZR1444600); the NYU Shanghai Boost Fund; the China Foreign Experts Program (G2021013002L); the NYU-ECNU Institute of Physics at NYU Shanghai; the NYU Shanghai Major-Grants Seed Fund; and Tamkeen under the NYU Abu Dhabi Research Institute grant CG008. 
\end{acknowledgments}

  \bibliographystyle{apsrev}
  \bibliography{references}

\appendix

\section{SINGLET STATES}

\subsection{General singlet states}

In this section we describe some basic properties of singlet states.  

A singlet state is defined as any quantum state satisfying
\begin{align}
    S^2\ket{\mathcal{S}} & = 0 \nonumber \\
    S^z \ket{\mathcal{S}}& = 0 .
\end{align}
The singlet state is invariant, up to a global phase, to all homogeneous qubit rotations 
\begin{align}
U^{\otimes N} | \mathcal{S} \rangle = e^{i \phi } | \mathcal{S} \rangle
\label{bilateral}
\end{align}
where $ U $ is an arbitrary single qubit unitary and $ \phi $ is a global phase. As such, the singlet state is unique among angular momentum eigenstates $ | s, \alpha, m \rangle $ in that it is 
rotationally symmetric, such that it is also a zero eigenstate of total spins $ S^j \ket{\mathcal{S}} = 0 $ in any direction $ j = \{x,y,z \} $.  

\begin{figure}[t]   
\includegraphics[width=\linewidth]{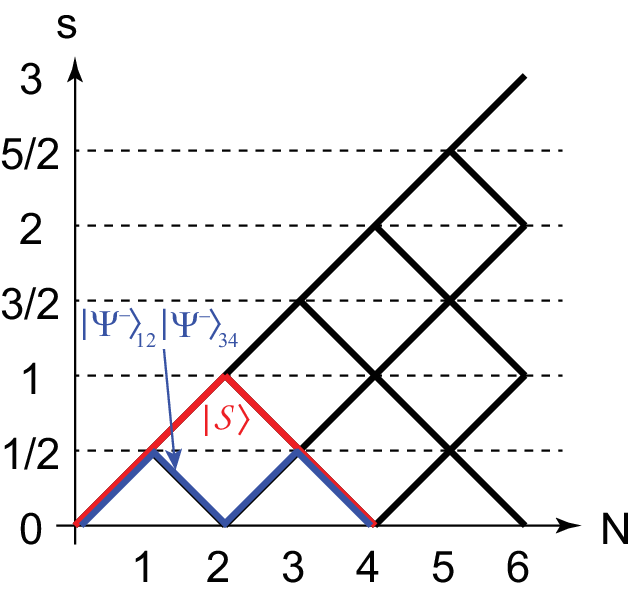}
\caption{Angular momentum coupling chart for $ N $ spin-$1/2$ (qubits).  Each path corresponds to a irreducible representation of the total angular momentum.}
\label{figa1}
\end{figure}

For $ N = 2 $ the singlet state is unique and coincides with the Bell state $ | \Psi^- \rangle = (|01 \rangle - | 10 \rangle )/\sqrt{2} $.  For larger (even) $ N $, the singlet state is not unique.  This can be understood in terms of angular momentum coupling.  
For example, coupling two spin $1/2$ gives the direct sum of two irreducible representations according to $ \tfrac{1}{2} \otimes \tfrac{1}{2} = 0 \oplus 1$.  Further performing tensor products gives angular momentum coupling according to Fig. \ref{figa1}.  At each point in the diagram there is a splitting since each tensor product with spin $ \tfrac{1}{2} $ can either increase or decrease the spin by $ \tfrac{1}{2} $ (with the exception of spin 0), such that each path gives a distinct coupling pattern.  Singlet states correspond to all paths that end with $ s = 0 $. 
The number of singlets for $ N $ qubits then corresponds to the number of paths that end with a total spin $ s = 0 $.  This means that the dimension of this subspace is the Catalan number \cite{edmonds1996angular}
\begin{align}
    C_{N/2} = \frac{1}{N/2+1} \binom{N}{N/2} .
\end{align}

Using the property (\ref{bilateral}), several general properties of singlet states can be deduced.  For single qubit expectation values we have
\begin{align}
\langle \mathcal{S} | X_n| \mathcal{S} \rangle &  =
\langle \mathcal{S} | (U^{\otimes N })^\dagger X_n  U^{\otimes N } | \mathcal{S} \rangle \nonumber \\
& = \langle \mathcal{S} | \sigma^j_n | \mathcal{S} \rangle  = 0, 
\label{singlequbitcor}
\end{align}
for $ j \in \{x,y,z \} $.  Such single qubit expectation values are zero because $ \langle \mathcal{S} | X_n| \mathcal{S} \rangle $ produces a single spin flip which removes the state from the $ m = 0 $ sector.  

Two qubit spin correlations along the same axis are symmetric:
\begin{align}
\langle \mathcal{S} | Z_n Z_m | \mathcal{S} \rangle &  =
\langle \mathcal{S} | (U^{\otimes N })^\dagger Z_n Z_m U^{\otimes N } | \mathcal{S} \rangle \nonumber \\
& = \langle \mathcal{S} | \sigma^j_n \sigma^j_m | \mathcal{S} \rangle \label{diagcor}
\end{align}
where $ \sigma^j $ are Pauli matrices along the $ j \in \{x,y,z \} $ axes and the $ U $ are chosen such as to rotate the Pauli matrices accordingly.  For correlations in different directions we have
\begin{align}
\langle \mathcal{S} | \sigma^i_n \sigma^j_m | \mathcal{S} \rangle  = 0 , \label{offdiagcor}
\end{align}
for $ i \ne j $, where we used the fact that $ \langle \mathcal{S} | X_n Z_m | \mathcal{S} \rangle = \langle \mathcal{S} | Z_n X_m | \mathcal{S} \rangle  = 0 $ since the $ X_n $ operators change the $ m=0 $ in a similar way to (\ref{singlequbitcor}).  The remaining correlations can be deduced using a similar argument to (\ref{diagcor}).  Finally, we can also deduce that 
\begin{align}
\langle \mathcal{S} | \sigma^+_n  \sigma^-_m | \mathcal{S}  \rangle = \frac{1}{2} \langle \mathcal{S} | Z_n Z_m | \mathcal{S} \rangle ,
\end{align}
where we used the identity
\begin{align}
\sigma^+_n\sigma^-_m = \frac{1}{4}\left( X_n X_m + Y_n Y_m +i(Y_n X_m-X_n Y_m)\right),
\end{align}
and the relations (\ref{diagcor}) and (\ref{offdiagcor}).

\subsection{Supersinglets}

For our multiparty QCS protocol, we are primarily interested in the supersinglet state \cite{cabello2003supersinglets}, as defined in (\ref{supersingletdef}).  Due to the basis invariance property (\ref{bilateral}), we may equally write the supersinglet as
\begin{align}
| {\cal S}_{N} \rangle =  & \sum_{m=-N/4}^{N/4}  \frac{(-1)^{N/4-m}}{\sqrt{N/2+1}} \nonumber \\
& \times |s=N/4,1,m \rangle_{N/2}^{(j)} | s=N/4,1,-m \rangle_{N/2}^{(j)}  .
\label{supersingletdefallbasis}
\end{align}
where $ j \in \{x,y,z \} $.   We see the supersinglets are formed by coupling each half of the $N$ qubits into a maximal $ s = N/4 $ spin, then coupling these two to form a spin zero state.  In Fig. \ref{figa1}, this corresponds to the upper triangular path that initially increases to $ s = N/4 $, then goes down to $ s = 0 $.  

The $ s = N/4 $ spin eigenstates can be written
\begin{align}
| D_{k= m + N/4}^{N/2} \rangle = |s=N/4,1,m \rangle^{(z)}_{N/2} 
\end{align}
where we defined the Dicke states on $ N/2 $ qubits with $ k $ excitations as
\begin{align}
& | D_{k}^{N/2} \rangle  =
\frac{1}{\sqrt{\binom{N/2}{k}}}\sum_{ \sigma } P_{\sigma} (
|0\rangle^{\otimes N/2-k} \otimes |1\rangle^{\otimes k} ),
\label{dickestate}
\end{align}
where $ P_\sigma ( \cdot)  $ is a permutation operator which for each $ \sigma $ produces a distinct rearrangement of the qubits.  For a state with $N/2 -k $ zeros and $ k $ ones, there are $ \binom{N/2}{k} $ such combinations. 

In terms of Dicke states, the supersinglet is (\ref{supersingletdef}), which can be further simplified to the form
\begin{align}
    \label{supersinglet2}
    \ket{\mathcal{S}_{N}} &  = \frac{1}{(N/2)!\sqrt{N/2+1}} \nonumber \\
   &  \times \sum_{\sigma}(-1)^{\hat{K}_1}\hat{K}_1!(N-\hat{K}_1)! P_\sigma (
   |0\rangle^{\otimes N/2}  \otimes |1\rangle^{\otimes N/2}  ) ,
\end{align}
where $\hat{K}_1 = S^z_1 + N/4 $ counts the number of ones in the first $N/2$ qubits, with $ S^z_1 = \sum_{n=1}^{N/2} \sigma^z_n/2 $.  In the above we used the identity
\begin{align}
& \sum_{k=0}^{N/2} \sum_{\sigma_1 \sigma_2} P_{\sigma_1} (  |0\rangle^{\otimes N/2-k} \otimes |1\rangle^{\otimes k}  ) P_{\sigma_2} (|0\rangle^{\otimes k}  \otimes |1\rangle^{\otimes N/2-k}  ) \nonumber\\
& = \sum_{\sigma} P_{\sigma} (  |0\rangle^{\otimes N/2} \otimes  |1\rangle^{\otimes N/2}   ) ,
\end{align}
which is similar in logic to the following combinatorial identity
\begin{align}
\binom{N}{N/2}=\sum_{k=0}^{N/2} \binom{N/2}{k}\binom{N/2}{N/2-k} .
\end{align}
Each register consisting of \( N \) qubits with exactly \( N/2 \) ones can be viewed as a subset of \( N/2 \) elements selected from a total set of \( N \) elements. One natural way to construct such a subset is by choosing \( k \) elements from the first \( N/2 \) qubits and \( N/2 - k \) elements from the second \( N/2 \) qubits. By allowing \( k \) to range from \( 0 \) to \( N/2 \), we generate all possible configurations of \( N/2 \) ones distributed across the \( N \) qubits.


\subsection{Homogeneous singlets}

Another type of singlet is the homogeneous singlet state, defined as
\begin{equation}
    \ket{\mathcal{S}_{\text{hom}}} = \frac{1}{\sqrt{\binom{N}{N/2}}} \sum_{\tau : |\tau| = N/2 } e^{i\omega_{\tau} } | \tau \rangle ,
    \label{homosinglet}
\end{equation}
where $  | \tau \rangle  $ is a computational basis state and the sum is restricted to states where $ \tau $ is a binary number with $ N/2 $ zeros and $ N/2 $ ones.  In order for the above state to be a singlet state 
\begin{equation}
    S^x\ket{\mathcal{S}_{\text{hom}}} = S^y \ket{\mathcal{S}_{\text{hom}}} = S^z\ket{\mathcal{S}
    _{\text{hom}}} = 0,
\end{equation}
the phases $ \omega_{\tau} $ must satisfy particular conditions.  The condition for the \( z \)-component is automatically satisfied, since the sum runs over the kernel of \( S_z \), i.e., the subspace with total magnetization zero.  
The \( x \)-component, however, yields a nontrivial constraint:
\begin{equation}
    \label{eq:Sx-action}
    \sum_{n=1}^{N} X_n \ket{S_{\text{hom}}}
    = \sum_{\tau : |\tau| = N/2 }  e^{i \omega_\tau} \sum_{n=1}^{N} X_n  | \tau \rangle =0,
\end{equation}
Projecting (\ref{eq:Sx-action}) onto any computational basis state \( | \tau \rangle \) with weight \(N/2\pm1\) yields the local consistency relations: for every \(\tau\) of weight \(N/2+1\),
\begin{equation}
\label{eq:constraint_plus}
    \sum_{n:\,\tau_n=1} e^{i\omega_{\tau^{(n)}}}=0,
\end{equation}
and for every \(\tau'\) of weight \(N/2-1\),
\begin{equation}
\label{eq:constraint_minus}
    \sum_{n:\,\tau'_n=0} e^{i\omega_{\tau'^{(n)}}}=0.
\end{equation}
Here \(\tau^{(n)}\) denotes the weight-\(N/2\) string obtained from \(\tau\) by flipping its \(n\)th bit. Equations (\ref{eq:constraint_plus}) and (\ref{eq:constraint_minus}) are equivalent statements (related by complement) and together are necessary and sufficient for \(S_x\ket{S_{\mathrm{hom}}}=0\). The \(S_y\) condition is not independent: since \(Y_n=i X_n Z_n\) and all \(\ket{\tau}\) lie in the \(S_z=0\) subspace, the \(S_y\) constraint imposes the vanishing of the same real and imaginary parts of the sums in (\ref{eq:constraint_plus})--(\ref{eq:constraint_minus}). Thus the singlet condition reduces to the requirement that, for every weight-\( N/2\pm1 \) string, the complex phases of its adjacent weight-\(N/2\) neighbors sum to zero. As an elementary illustration, for \(N=4\) the weight-3 string \(\tau=1110\) produces the constraint \(e^{i\omega_{0110}}+e^{i\omega_{1010}}+e^{i\omega_{1100}}=0\), and analogous relations hold for all weight-3 and weight-1 strings.

In this state, the bipartite entanglement between two parties, measured by concurrence, is always zero \cite{LONE201573}.  However, entanglement between a pair and the remaining parties is maximized in such a state.




\section{QCS SIGNAL}

In the QCS protocol, Alice performs the projective measurement $  \Pi_\pm = |\pm \rangle \langle \pm | $.  We consider the two measurement outcomes by Alice separately.  In the case that Alice obtains the outcome $ | + \rangle$, the state is
\begin{align}
|\psi(t) \rangle & = \frac{1}{\sqrt{p_+}} e^{-iHt/\hbar} \Pi_+ | \mathcal{S} \rangle  \nonumber \\
& =  e^{-iHt/\hbar} |+ \rangle_1 |s=1/2,m=-1/2 \rangle_{N-1}^{(x)} ,
\label{conditionalstate}
\end{align}
Here $ p_\pm = 1/2 $ are the probabilities of the two outcomes. Meanwhile, if Alice obtains the outcome $ | - \rangle$, the state is
\begin{align}
|\psi (t) \rangle & =  \frac{1}{\sqrt{p_+}} e^{-iHt/\hbar} e^{i\pi S^z}  \Pi_- | \mathcal{S} \rangle  \nonumber \\
& =  e^{-iHt/\hbar} e^{i\pi S^z}  | -\rangle_1 |s=1/2,m=1/2 \rangle_{N-1}^{(x)} , \nonumber \\
& =  e^{-iHt/\hbar}  | + \rangle_1 |s=1/2, m= -1/2 \rangle_{N-1}^{(x)} 
\end{align}
where we used the fact that a $ \pi$ rotation of $ S^z = Z_1/2 + \sum_{n=2}^{N} Z_n/2 $ is performed for this case. This shows that for either measurement outcome we have the same state (\ref{postmeasuredpsi}). The signal (\ref{qcsamp}) is then evaluated in the Heisenberg picture 
\begin{align}
 e^{iHt/\hbar } X_n e^{-iHt/\hbar } = X_n  \cos \omega t -Y_n  \sin \omega t 
 \label{heisenbergeval}
\end{align}
and we used $\langle \psi_+ (0) | Y_n | \psi_+ (0) \rangle = 0$.

Alternatively, one could also classically post-process the expectation values, instead of performing the physical correction $e^{i\pi S^z}$. In this case, we have the two states
\begin{align}
|\psi_+(t) \rangle & = \frac{1}{\sqrt{p_+}} e^{-iHt/\hbar}  \Pi_+ | \mathcal{S} \rangle  \nonumber \\
& =  e^{-iHt/\hbar}  |+ \rangle_1 |s=1/2, l,m=-1/2 \rangle_{N-1}^{(x)} \label{psiplus} 
\end{align}
and
\begin{align}
|\psi_- (t) \rangle & =  \frac{1}{\sqrt{p_+}} e^{-iHt/\hbar} \Pi_- | \mathcal{S} \rangle  \nonumber \\
& =  e^{-iHt/\hbar} | -\rangle_1 |s=1/2, l,m=1/2 \rangle_{N-1}^{(x)} .
\end{align}

Evaluating the expectation value for Alice's $ | + \rangle $ outcome using (\ref{psiplus}), we have
\begin{align}
\langle \psi_+ (t) | X_n | \psi_+ (t) \rangle = 
 \frac{1}{p_+}   \langle \mathcal{S} | \Pi_+  X_n \Pi_+  | \mathcal{S}   \rangle \cos \omega t
\end{align}
where we evaluated the time evolution in the Heisenberg picture (\ref{heisenbergeval}). For the other outcome, we have similarly
\begin{align}
\langle \psi_-(t) | X_n | \psi_- (t) \rangle = 
 \frac{1}{p_-}  \langle \mathcal{S} | \Pi_-  X_n \Pi_-  | \mathcal{S}   \rangle \cos \omega t
\end{align}
Since the expectation $ \Pi_+  | \mathcal{S}   \rangle \propto  |s=1/2, l,m=-1/2 \rangle_{2N-1}^{(x)} $ and $ \Pi_- | \mathcal{S}   \rangle \propto  |s=1/2, l,m=1/2 \rangle_{2N-1}^{(x)} $, simply performing an average gives zero signal.  By modifying the signal for Alice's $ | - \rangle $ outcomes by a factor of $-1$, then taking the average, we obtain a non-zero signal.  Specifically, the signal from the $n$th party is
\begin{align}
f_n(t) & = p_+ \langle \psi_+ (t) | X_n | \psi_+ (t) \rangle 
- p_-\langle \psi_-(t) | X_n | \psi_- (t) \rangle \nonumber \\
& =  \left( \langle \mathcal{S} | \Pi_+  X_n \Pi_+  | \mathcal{S}   \rangle - 
 \langle \mathcal{S} | \Pi_-  X_n \Pi_-  | \mathcal{S}   \rangle \right) \cos \omega t
 \label{signalclassicalcor}
\end{align}
which gives the same signal as (\ref{qcsamp}).  

The right hand side of (\ref{signalclassicalcor}) can be manipulated into another form
\begin{align}
f_n(t) = \langle \mathcal{S} | X_1 X_n | \mathcal{S} \rangle  \cos \omega t
\label{signalxx}
\end{align}
where we used the fact that $ \Pi_\pm^2 = \Pi_\pm $, $ [ \Pi_\pm,  X_n ] = 0  $, and $ X_1 =  \Pi_+ -  \Pi_- $.  Thus the amplitude of the signal is related to the two-qubit correlation of the singlet state.  Using the general property of singlets (\ref{diagcor}) we can equally write
\begin{align}
f_n(t) =  \langle \mathcal{S} | Z_1 Z_n | \mathcal{S} \rangle  \cos \omega t .
\label{signalzzform}
\end{align}
This means that the accuracy of a member's signal is directly related to their qubit's correlation with Alice.

The corresponding relation to (\ref{totalamp}) can be obtained using the relation
\begin{align}
Z_1 S^z | \mathcal{S} \rangle & = \frac{1}{2} Z_1 ( Z_1 + \sum_{n=2}^{N} Z_n ) | \mathcal{S} \rangle \nonumber \\
& = \frac{1}{2} ( I  + Z_1 \sum_{n=2}^{N} Z_n ) | \mathcal{S} \rangle = 0 ,
\end{align}
since $ S^z | \mathcal{S} \rangle = 0 $. Taking the inner product with $ \langle \mathcal{S}  | $ we obtain
\begin{equation}
   \sum_{n=2}^{N}  \langle \mathcal{S}  | Z_1 Z_n  | \mathcal{S} \rangle  = -1 .
    \label{sumsigs}
\end{equation}
%

\section{QCS AMPLITUDE EVALUATION}

\subsection{Correlation method}

Here we evaluate the amplitudes (\ref{supersingletamplitudes}) of the QCS protocol using the supersinglet state.  Using the form of the signal (\ref{signalzzform}), we evaluate 
\begin{align}
\langle \mathcal{S}  | Z_1 Z_n  | \mathcal{S} \rangle & = 
p_{00}^{(1,n)} + p_{11}^{(1,n)} - p_{01}^{(1,n)} -p_{10}^{(1,n)}  \\
& = 4 p_{00}^{(1,n)} - 1
\label{amplitudezz}
\end{align}
where we defined the probabilities for $ l_n \in \{0,1 \} $
\begin{align}
p_{l_1 l_n }^{(1,n)} = |\langle l_1 l_n | \mathcal{S}_{} \rangle |^2
\end{align}
and we used the fact that $ p_{00}^{(1,n)} + p_{11}^{(1,n)} + p_{01}^{(1,n)} + p_{10}^{(1,n)}  = 1$ and also the symmetry of the Clebsch-Gordan coefficients $ \forall l \in\{0,1\}^{N}$ for singlet states, $\bra{l}\ket{\cal S}=(-1)^{N/2} \bra{\bar{l}}\ket{\cal S} $, where $\bar{l}$ is the conjugate of the bitstring $l$. 

Considering the case $ n \in [2,N/2] $, we evaluate
\begin{align}
p_{00}^{(1,n)} = \sum_{\bar{k}=2}^{N/2} p_{00| \bar{k} }^{(1,n)} p_{\bar{k}}
\label{probsum1}
\end{align}
where
\begin{align}
    p_{\bar{k}} = \frac{1}{N/2+1}
\end{align}
is the probability that the supersinglet state contains $ \bar{k} $ zeros in the first $ N/2 $ bits, corresponding to $ k = N/2 - \bar{k} $ in (\ref{supersingletdef}).  Meanwhile,
\begin{align}
p_{00| \bar{k} }^{(1,n)} = \frac{\binom{N/2-2}{\bar{k}-2}\binom{N/2}{N/2-\bar{k}}}{\binom{N/2}{\bar{k}}\binom{N/2}{N/2-\bar{k}}} = \frac{2k(k-1)}{N(N/2-1)}
\end{align}
is the probability of obtaining the outcome \( |00\rangle  \) when measuring the first (Alice's) qubit and the $ n$th qubit, assuming the first $N/2$ qubits are in the Dicke state \( \left| k = N/2 - \bar{k} \right\rangle_N \), which contains \( \bar{k} \) zeros. This is computed by placing \( \bar{k} - 2 \) zeros among the remaining \( N/2-2 \) qubits (since the two zeros corresponding to the first and $n$th qubit are already fixed), multiplied by the degeneracy of the Dicke state over the last \( N/2 \) qubits.

Evaluating (\ref{probsum1}), we obtain
\begin{align}
p_{00}^{(1,n)} = \frac{2}{N(N/2+1)(N/2-1)} \sum_{\bar{k}=2}^{N/2} \bar{k}(\bar{k}-1) = \frac{1}{3}    .  
\label{p00groupI}
\end{align}
Substituting this into (\ref{amplitudezz}), we have
\begin{align}
\langle \mathcal{S}  | Z_1 Z_n  | \mathcal{S} \rangle & =\frac{1}{3} . 
\end{align}

Next, for the case $ n \in [N/2+1,N] $, we have
\begin{align}
p_{00}^{(1,n)} = \sum_{\bar{k}=1}^{N/2} p_{00| \bar{k} }^{(1,n)} p_{\bar{k}} ,
\label{probsum2}
\end{align}
where the probability of obtaining $ |00\rangle  $ in the state $ | k = N/2 - \bar{k} \rangle_{N/2} |\bar{k} \rangle_{N/2}  $ is 
\begin{align}
p_{00| \bar{k} }^{(1,n)} =   \frac{\binom{N/2-1}{\bar{k}-1} \binom{N/2-1}{N/2-\bar{k}-1} }{\binom{N/2}{\bar{k}}\binom{N/2}{N/2-\bar{k}}}=\frac{4 \bar{k}(N/2-\bar{k})}{N^2} ,
\end{align}
where we place \( \bar{k} - 1 \) zeros among the first \( N/2 - 1 \) qubits, excluding Alice's qubit (since hers is already fixed to 0); and similarly place \( N/2 - \bar{k} - 1 \) ones among the last \( N/2 - 1 \) qubits, excluding the $n$th qubit (also fixed to 0). 

Evaluating (\ref{probsum2}), we obtain
\begin{align}
p_{00}^{(1,n)} = \frac{4}{N^2(N/2+1)} \sum_{\bar{k}=1}^{N/2} \bar{k}(N/2-\bar{k}) = \frac{N/2-1}{3N} .
\label{p00groupII}
\end{align}
Substituting this into (\ref{amplitudezz}), we have
\begin{align}
\langle \mathcal{S}  | Z_1 Z_n  | \mathcal{S} \rangle & = -\frac{N+4}{3N} . 
\end{align}

\subsection{Wavefunction method}

In this section we calculate the amplitudes (\ref{supersingletamplitudes}) using a direct wavefunction method.  
For the purposes of this section, we rotate Alice's measurement basis from $ x $ to $ z $ for notational convenience. Since singlet states are basis invariant, we use the supersinglet written in the form (\ref{supersingletdef}) written in terms of Dicke states (in the $ z $ basis).  

Written in the form (\ref{supersingletdef}), it is apparent that the qubits fall into two groups.  Applying a projection operator $ \Pi_l = | l \rangle_1 \langle l |_1 $ for $ l \in \{0,1\} $ on the first qubit, we obtain
\begin{align}
\Pi_0 |D_k^{N/2} \rangle  & =\frac{1}{\sqrt{\binom{N/2}{k}}} | 0 \rangle \sum_{ \sigma } P_\sigma (
| |0 \rangle^{\otimes N-k-1} \otimes |1 \rangle^{\otimes k} \rangle  )\nonumber \\
& =\sqrt{\frac{N-2k}{N}} | 0 \rangle |D_k^{N/2-1} \rangle  .
\end{align}
Similarly, we have 
\begin{align}
\Pi_1  |D_k^{N/2} \rangle  & =\sqrt{\frac{2k}{N}} | 1 \rangle |D_{k-1}^{N/2-1} \rangle.
\end{align}

Applying these to to the singlet state we have
\begin{align}
| \psi_0 \rangle  = & \frac{\Pi_0 |{\cal S} \rangle}{\sqrt{p_0}} \nonumber \\
= &  \sqrt{\frac{4}{N+2}} \nonumber \\
& \times \sum_{k=0}^{N/2-1} (-1)^{N/2-k} \sqrt{\frac{N-2k}{N}}
|D_k^{N/2-1} \rangle  |D_{N/2-k}^{N/2} \rangle
\end{align}
and
\begin{align}
| \psi_1 \rangle & = \frac{\Pi_1 |{\cal S} \rangle}{\sqrt{p_1}} \nonumber \\
& = \sqrt{\frac{4}{N+2}} \sum_{k=1}^{N/2} (-1)^{N/2-k} \sqrt{\frac{2k}{N}}
|D_{k-1}^{N/2-1} \rangle  |D_{N/2-k}^{N/2} \rangle .
\end{align}
The probabilities of the two outcomes $l \in \{0,1 \} $ are
\begin{align}
p_l = \langle {\cal S} | \Pi_l  \Pi_l |  {\cal S} \rangle = \frac{1}{2} . 
\end{align}


Evaluating the expectation values for the $ l = 0 $ outcome, we have for $ n \in [2,N/2] $ in the first group of qubits
\begin{align}
& \langle \psi_0 | Z_n | \psi_0 \rangle  = \frac{2 \langle \psi_0 | S^z_1| \psi_0 \rangle }{N/2-1} \nonumber \\
& = \frac{4}{N(N+2)} \sum_{k=0}^{N/2-1} (N-2k) \left(\frac{(N/2-1)-2k}{N/2-1} \right) \nonumber \\
& = \frac{1}{3}  ,
\label{1o3result}
\end{align}
where $S^z_1 = \sum_{n=2}^{N/2} Z_n/2  $ is the total spin operator for the first group.  For $ n \in [N/2+1, N] $ in the second group of qubits we evaluate 
\begin{align}
\langle \psi_0 | Z_n | \psi_0 \rangle & = \frac{2 \langle \psi_0 | S^z_2| \psi_0 \rangle }{N/2} \nonumber \\
& = \frac{4}{N(N+2)} \sum_{k=0}^{N/2-1} (N-2k) \left(\frac{2k-N/2}{N/2} \right)  \nonumber \\
& = -\frac{N+4}{3N} ,
\end{align}
where $S^z_2 = \sum_{n=N/2+1}^{N} Z_n /2 $.  In the above, we used the fact that the qubits within each Dicke state is completely symmetric such that $ \langle Z_n \rangle = \langle S^z \rangle/N$.  

For the $l =1 $ outcome, in the first group of qubits $ n \in [2,N/2] $
\begin{align}
\langle \psi_1 | Z_n | \psi_1 \rangle & = \frac{2\langle \psi_1 | S^z_1| \psi_1 \rangle }{N/2-1} \nonumber \\
& = \frac{8}{N(N+2)} \sum_{k=1}^{N/2} k \left(\frac{(N/2-1)-2(k-1)}{N/2-1} \right)  \nonumber \\
& = -\frac{1}{3} .
\end{align}
For $ n \in [N/2+1, N] $ in the second group of qubits,
\begin{align}
\langle \psi_1 | Z_n | \psi_1 \rangle & = \frac{\langle \psi_1 | S^z_2| \psi_1 \rangle }{N} \nonumber \\
& = \frac{8}{N(N+2)} \sum_{k=1}^{N/2} k \left(\frac{2k-N/2}{N/2} \right) \nonumber \\
& = \frac{N+4}{3N} .
\end{align}

Using the classical post-processing approach  (\ref{signalclassicalcor}) , we obtain a signal
\begin{align}
f_n(t) = \left\{
\begin{array}{cc}
\frac{1}{3} \cos \omega t  & n \in [2,N/2] \\
- \frac{N+4}{3N} \cos \omega t  & n \in [N/2+1,N] \\
\end{array}
\right. .
\end{align}

\subsection{Other singlet states}

We find that other choices of singlet states have a smaller magnitude of oscillations.  
Formally, we state our problem as to find a spin zero state 
\begin{align}
|{\cal S} \rangle = \sum_\alpha \phi_\alpha |s=0,\alpha, m=0 \rangle
\end{align}
such that the total signal amplitude
\begin{align}
{\cal A} = \left( \prod_{n=2}^{N} |A_n| \right)^{\tfrac{1}{N-1}}
\end{align}
is maximized.  We take the absolute value of the amplitude since we wish to obtain a large signal irrespective of sign.  The geometric mean favors equal amplitudes for all parties.

For example, another choice of singlet state is the homogenous singlet (\ref{homosinglet}) where all parties have the same amplitude.  The homogeneous singlet state yields the same results for all parties:
\begin{equation}
\langle \psi_0 | Z_n | \psi_0 \rangle = -  \frac{1}{N -1}, 
\end{equation}
which satisfies (\ref{totalamp}).  This has an amplitude that is lower than the supersinglet for $ N \ge 4 $.  For $ N > 4 $ it is evident that $ 1/(N-1) < 1/3 $, comparing to (\ref{1o3result}).  For $ N = 4 $ from Fig. \ref{fig2}(a) we saw that the supersinglet is optimal.

\section{OPTIMALITY OF THE SUPERSINGLET STATE}

In this section we show that the supersinglet state is the optimal spin zero state in terms of obtaining the maximum signal for all parties.

First, let us make some general statements about the range that the expectation value $ \langle {\cal S} | Z_1 Z_n |  {\cal S} \rangle $ can take.  We saw from (\ref{amplitudezz}) that these are directly related to the QCS signal amplitudes $ A_n$.  Since the spin zero state $ | {\cal S} \rangle $ is SU(2) invariant, when reduced to qubits 1 and $ n $, the state can only take the specific form
\begin{align}
\rho_{1n} & = x | \Psi^- \rangle \langle \Psi^-| + (1-x) \frac{I}{4} \label{2qubitrho} \\
& = \frac{1+3x}{4} | 0,1,0\rangle \langle  0,1,0 | \nonumber   \\
& + \frac{1-x}{4} \sum_{m=-1}^1 
| 1,1,m \rangle \langle  1,1,m | ,
\label{Mqubitrho}
\end{align}
since the only SU(2) invariant states for two qubits are the singlet state $ | \Psi^-\rangle $ and the completely mixed state $ I/ 4 $. In the second equality we wrote the states in terms of the two qubit singlet state and triplet states.  Demanding that the eigenvalues of $ \rho_{1n} $ are non-negative, we obtain the range of validity of the parameter $ -1/3 \le x \le 1 $.  We see that for $ x = -1/3 $, the density matrix is an equal mixture of the $ s = 1 $ triplet, and for $ x = 1 $ the state is purely the $ s = 0 $ singlet.  

The expectation for the state (\ref{2qubitrho}) may be evaluated as 
\begin{align}
\langle Z_1 Z_n \rangle = -x .  
\end{align}
It follows that the amplitude is bounded by region
\begin{align}
-1 \le A_n \le \frac{1}{3}
\end{align}
We must however simultaneously satisfy the constraint (\ref{totalamp}).  If we take all amplitudes to be the same value, we recover the homogenous singlet result
\begin{align}
A_n = -\frac{1}{N-1} .
\end{align}
However, it is evident that one can make a larger magnitude amplitude under constraint (\ref{totalamp}) by making as many of the amplitudes $ A_n = 1/3 $ as possible.  The remaining then can be chosen negative such that constraint (\ref{totalamp}) is satisfied. 

Following this logic, now suppose we choose $ M $ of the $N $ qubits to have a signal $ A_n = 1/3 $.  The physical state for these $ M $ qubits can be obtained by generalizing (\ref{Mqubitrho}) as 
\begin{align}
\rho_{12 \dots M} & = \frac{1}{M+1} \sum_{m=-M/2}^{M/2} | \tfrac{M}{2}, 1 , m 
\rangle \langle  \tfrac{M}{2},1, m  | \\
& = \frac{1}{M+1} \sum_{k=0}^{M} | D_k^M \rangle  \langle  D_k^M  |
\label{mstateferro}
\end{align}
The $ ZZ $ expectation values for this state can be evaluated to be
\begin{align}
\langle Z_1 Z_n \rangle = A_n = \frac{1}{3} 
\end{align}
for $ n \in [2,M] $. 

The maximum spin of the $M $ qubits (i.e. $ s = M/2 $) creates the largest amplitudes $A_n$. If instead we consider the state 
\begin{align}
\rho_{12 \dots M}^{(s)} = \frac{1}{2s+1} \sum_{m=-s}^{s} | s, \alpha , m 
\rangle \langle  s, \alpha, m  | ,
\label{mstateferros}
\end{align}
the $ ZZ$ expectation values take a value
\begin{align}
\langle Z_1 Z_n \rangle = A_n = \frac{4s(s+1) - 3N}{3N(N-1)}
\end{align}
which takes a maximum value for $ s = N/2 $.  Hence in the interest of making the largest values of $ A_n$, it is beneficial to choose the largest spin $ s = M/2 $.  

Now our problem reduces to finding a spin zero state with the reduced density matrix (\ref{mstateferro}), where we would like to take $ M $ as large as possible.  Since (\ref{mstateferro}) is an $ s = M/2 $ state, the possible range of $ M $ is $ 0 \le M \le N/2 $, since for more than $ M = N/2 $ it is impossible to form a spin zero state with the remaining $ N - M $ qubits.  Taking the largest $ M = N/2 $, we see that the supersinglet has exactly the reduced density matrix (\ref{mstateferro}), since it couples two $ s = N/4 $ spins with equal Dicke state amplitudes.

\section{DECOHERENCE}


To evaluate the effects of decoherence on the QCS protocol, we consider a model of dephasing where there are independent phase-flip errors that occur according to 
\begin{equation}
\mathcal{E}(\rho)
= \sum_{s\in\{0,1\}^{N}} p^{|s|}(1-p)^{N-|s|}\; Z^{s}\,\rho\,Z^{s},
\label{dephasingchannel}
\end{equation}
where $\quad
Z^{s}=\bigotimes_{j=1}^{N} Z^{s_j}\ (\,Z^0\!=\!I,\ Z^1\!=\!Z\,)$. The above channel corresponds to each qubit having a probability of having a phase-flip error with probability $ p $.  The binary number $ s $ specifies the type of error on the state, from no errors $ s = 00 \dots 0 $ to errors occurring on all qubits $ s = 11 \dots 1 $.  The probability of each outcome is calculated according to how many errors $ |s | $ occur.  

\begin{figure}[t]   
\includegraphics[width=\linewidth]{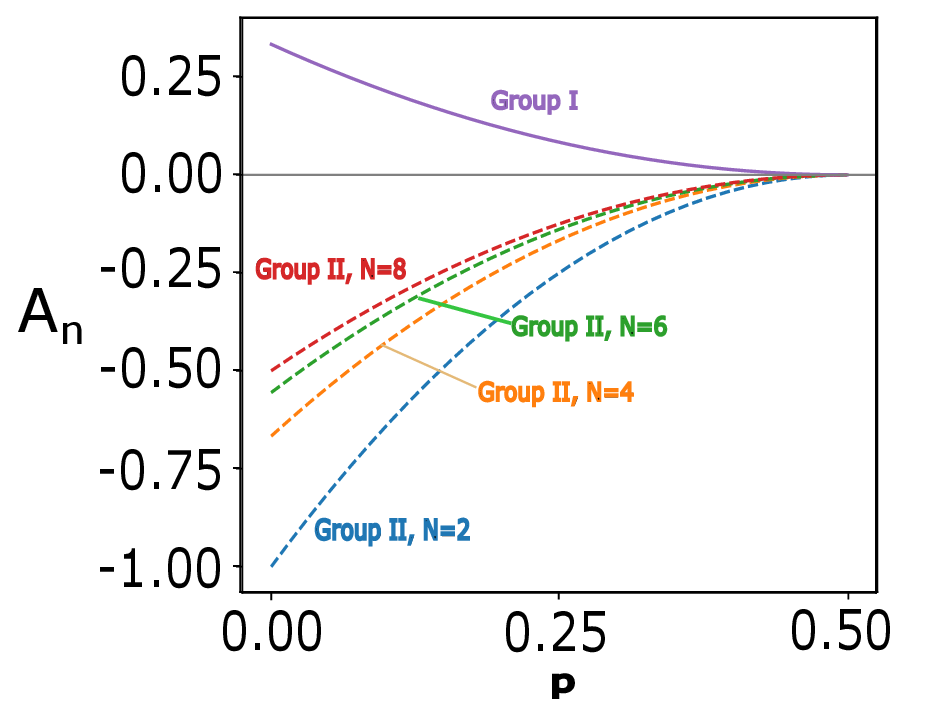}
\caption{Signal amplitudes $ A_n = \text{Tr}( \rho  X_1 X_n ) $  after the dephasing channel (\ref{dephasingchannel}) as a function of $p$.
    The solid curve show the signal of parties in Group I ($n \in [2,N] $), which remains 
    invariant with respect to the total number of parties $N$. The dashed curves 
     represent the signal of Group II ($n \in [N/2+1,N] $), for different system sizes $N$.}
\label{figa2}
\end{figure}

Assuming the noise acts after state preparation, the resulting QCS signal is modified accordingly, allowing us to analyze the degradation of the signal amplitude (Fig. \ref{figa2}).  We see that as expected the signal degrades with the presence of the dephasing channel, reaching zero signal quadratically for $ p = 1/2 $.

\section{ERROR ANALYSIS}

Here we evaluate the accuracy of our supersinglet QCS protocol.  We follow the methods given in Ref. \cite{ilo2018remote}.  One source of error in the QCS protocol is an imperfect fidelity of the supersinglet state, which may occur to insufficient rounds of purification, for example.  The worst type of error that can occur is a systematic $ Z$ rotation of the qubits, where Alice's qubit is rotated in the opposite direction to the remaining qubits
\begin{align}
| {\cal S}' \rangle & = e^{-i \epsilon Z_1/2} e^{i \epsilon \sum_{n=2}^N Z_n/2}  | {\cal S} \rangle \nonumber \\
& = e^{i \epsilon \sum_{n=2}^N Z_n}  | {\cal S} \rangle = e^{-i \epsilon Z_1} | {\cal S} \rangle ,
\end{align}
where in the second line we applied a factor of $ I =  \exp (i \epsilon S^z) \exp (-i \epsilon S^z )$ and used the fact that $ \exp (\pm i \epsilon S^z)  | {\cal S} \rangle =  | {\cal S} \rangle $.  The fidelity of this state compared to the original state is 
\begin{align}
F  & = | \langle {\cal S} | {\cal S}' \rangle |^2 =  \cos^2 \epsilon , 
\label{fiderror}
\end{align}
where we used the form (\ref{belllike}).  

The state (\ref{conditionalstate}) in the case that the error is included becomes
\begin{align}
|\psi'(t) \rangle = e^{-i (\omega t/2 -\epsilon) \sum_{n=2}^N Z_n} | s=1/2, m= -1/2 \rangle_{N-1}^{(x)} .
\label{errorpsi}
\end{align}
Hence the presence of a remnant Preskill phase simply changes $ \omega t \rightarrow \omega t - 2 \epsilon $.  Following the steps leading to (\ref{signalzzform}), the QCS signal is then modified as
\begin{align}
f_n (t) = A_n \cos (\omega t - 2\epsilon) .
\label{offsetsignal}
\end{align}
We see that a remnant Preskill phase gives a systematic time offset of 
\begin{align}
  \delta t_F = \frac{2 \epsilon }{\omega} \approx \frac{2 \sqrt{1-F}}{\omega}
  \label{deltaferror}
\end{align}
where we used (\ref{fiderror}) to convert the error $ \epsilon $ to a fidelity. In using (\ref{fiderror}), we assume a worst-case state in the context of QCS that gives the largest error in the timing signal for a given fidelity $ F $. 

Another source of error is due to statistical noise in estimating the QCS signal.  To estimate this consider the measurement on the $ n $th qubit.  Consider the measurement outcome $ |+ \rangle $ of the state (\ref{errorpsi}).  The probability of obtaining 
$ |  + \rangle $  on the $ n $th qubit is 
\begin{align}
p_+ & =  \langle \psi'(t)|  + \rangle  \langle +  | \psi'(t) \rangle \nonumber \\
& =  \langle s=1/2, m= -1/2 |^{(x)} \Big[ \cos^2 (\frac{\omega t - 2 \epsilon}{2} ) | + \rangle\langle + |   \nonumber \\ 
& - i \sin (\frac{\omega t - 2 \epsilon}{2} ) \cos (\frac{\omega t - 2 \epsilon}{2} ) | - \rangle \langle + |  \nonumber \\
& + i \sin (\frac{\omega t - 2 \epsilon}{2} ) \cos (\frac{\omega t - 2 \epsilon}{2} ) | + \rangle \langle - |  \nonumber \\
& + \sin^2 (\frac{\omega t - 2 \epsilon}{2} ) | -\rangle\langle - |
\Big] | s=1/2, m= -1/2  \rangle^{(x)}  \nonumber \\
& = 2 p_{00}^{(1,n)} \cos^2 (\frac{\omega t - 2 \epsilon}{2} ) + 
2 p_{01}^{(1,n)}  \sin^2 (\frac{\omega t - 2 \epsilon}{2} ) . 
\end{align}
Here, the probability $ p_{00}^{(1,n)} $ was evaluated in (\ref{p00groupI}) and (\ref{p00groupII}) for $ n \in [2,N/2] $ and $ n \in [N/2+1,N] $ respectively.  From the fact that $ p_{00}^{(1,n)}=p_{11}^{(1,n)}$ and $ p_{01}^{(1,n)}=p_{10}^{(1,n)}$ and the probabilities sum to 1 we can deduce
\begin{align}
p_{01}^{(1,n)}= \frac{1}{2} -   p_{00}^{1,n} .
\end{align}
The probability of the outcome $ |-  \rangle $ is
\begin{align}
p_- & =  \langle \psi'(t)| - \rangle  \langle -  | \psi'(t) \rangle \nonumber \\
& =  \langle s=1/2, m= -1/2 |^{(x)} \Big[ \cos^2 (\frac{\omega t - 2 \epsilon}{2} ) |- \rangle\langle - |   \nonumber \\ 
& - i \sin (\frac{\omega t - 2 \epsilon}{2} ) \cos (\frac{\omega t - 2 \epsilon}{2} ) | - \rangle \langle + |  \nonumber \\
& + i \sin (\frac{\omega t - 2 \epsilon}{2} ) \cos (\frac{\omega t - 2 \epsilon}{2} ) | + \rangle \langle - |  \nonumber \\
& + \sin^2 (\frac{\omega t - 2 \epsilon}{2} ) | +\rangle\langle + |
\Big] | s=1/2, m= -1/2  \rangle^{(x)}  \nonumber \\
& =2 p_{01}^{(1,n)}   \cos^2 (\frac{\omega t - 2 \epsilon}{2} ) + 
 2 p_{00}^{(1,n)} \sin^2 (\frac{\omega t - 2 \epsilon}{2} )  .
\end{align}

To estimate the signal (\ref{offsetsignal}), the $ n$th party makes $ M $ measurements.  For a particular set of $ M $ measurements, the probability of obtaining $ k $ outcomes with $ | + \rangle $ is 
\begin{align}
p_k & = \binom{M}{k}  p_+^k p_-^{M-k} \nonumber \\
& \approx \frac{1}{\sqrt{2 \pi M p_+ p_-}} \exp 
\Big[ - \frac{M}{4 p_+ p_-} \big( \frac{2k-M}{M} - (p_+ - p_-) \big)^2 \Big] , 
\end{align}
where in the second line we used a Gaussian approximation \cite{Ilo-okeke2015} to approximate the distribution.  The variable $ x = \frac{2k-M}{M} $ takes the range $ [-1,1] $ and has a peak at the value
\begin{align}
p_+ - p_- =  2 ( p_{00}^{(1,n)} - p_{01}^{(1,n)} ) \cos (\omega t - 2 \epsilon ),
\end{align}
which agrees with (\ref{offsetsignal}).  The standard deviation of the variable $ x= \frac{2k-M}{M} $ is
\begin{align}
\delta x \approx \sqrt{ \frac{2 p_+ p_-}{M}}  .
\end{align}
Taking the average value of $ p_{\pm} $ over a period, we have
\begin{align}
\bar{p}_{\pm} = p_{00}^{(1,n)} + p_{01}^{(1,n)} = \frac{1}{2} .
\end{align}

\newpage

Thus the typical standard variation of $ \delta x \approx 1/\sqrt{2M} $, which equates to a time variance of
\begin{align}
\delta t_{\text{SQL}} \approx \frac{1}{\omega \sqrt{2M}} .
\label{sqlerror}
\end{align}
Combining the errors due to the phase offset and the statistical noise we have
\begin{align}
\delta t & \approx \sqrt{ \delta t_F^2 + \delta t_{\text{SQL}}^2 } 
\end{align}
which leads to (\ref{totalerror}) on substituting (\ref{sqlerror}) and (\ref{deltaferror}).

\end{document}